\documentclass[twocolumn,showpacs,preprintnumbers,amsmath,amssymb,prl,superscriptaddress,longbibliography]{revtex4-1}
\usepackage{bbm}
\usepackage{mathrsfs}
\usepackage{graphicx}
\usepackage{dcolumn}
\usepackage{bm}
\usepackage{amsmath}
\usepackage{amsfonts}
\usepackage{color}
\usepackage[colorlinks=true,linkcolor=magenta,urlcolor=magenta,citecolor=cyan,anchorcolor=blue]{hyperref}
\usepackage{floatrow}

\begin{document}

\title{Large-Gap Quantum Anomalous Hall Insulators in \textit{A}Ti\textit{X} Class}
\author{Yadong Jiang}
\affiliation{State Key Laboratory of Surface Physics and Department of Physics, Fudan University, Shanghai 200433, China}
\author{Huan Wang}
\affiliation{State Key Laboratory of Surface Physics and Department of Physics, Fudan University, Shanghai 200433, China}
\author{Jing Wang}
\thanks{Corresponding author: wjingphys@fudan.edu.cn}
\affiliation{State Key Laboratory of Surface Physics and Department of Physics, Fudan University, Shanghai 200433, China}
\affiliation{Institute for Nanoelectronic Devices and Quantum Computing, Fudan University, Shanghai 200433, China}
\affiliation{Zhangjiang Fudan International Innovation Center, Fudan University, Shanghai 201210, China}

\begin{abstract}
We theoretically propose that the monolayer $A$Ti$X$ family (KTiSb, KTiBi, RbTiSb, SrTiSn) are potential candidates for large-gap quantum anomalous Hall insulators with high Chern number $\mathcal{C}=2$. Both of the topology and magnetism in these materials are from $3d$-orbitals of Ti. We construct the tight-binding model with symmetry analysis to reveal the origin of topology. Remarkably, quite different from the conventional $s$-$d$ band inversion, here the topological band inversion within $3d$ orbitals is due to the crystal field and electron hopping, while spin-orbit coupling only trivially gaps out the Dirac cone at Fermi level. The general physics from the $3d$ orbitals here applies to a large class of transition metal compounds with the space group $P4/nmm$ or $P$-$42m$ and their subgroups.
\end{abstract}

\date{\today}


\maketitle

{\color{blue}\emph{Introduction.}} 
The precise theoretical prediction and experimental realization of the quantum anomalous Hall (QAH) effect set a remarkable example for understanding and engineering topological states of quantum matter in complex materials~\cite{hasan2010,qi2011,tokura2019,wang2017c,chang2022,bernevig2022}. The QAH insulator has a topologically nontrivial electronic structure with a finite Chern number~\cite{thouless1982} characterized by a bulk energy gap but gapless chiral edge states, leading to the quantized Hall effect without an external magnetic field~\cite{haldane1988}. The QAH effect has been realized in magnetically doped topological insulators (TI)~\cite{chang2013b,chang2015,mogi2015,bestwick2015,watanabe2019}, in the intrinsic magnetic TI MnBi$_2$Te$_4$~\cite{deng2020}, in the twisted bilayer graphene~\cite{serlin2020} and transition metal dichalcogenide heterobilayers~\cite{li2021}, with comparable onset temperature of about a few Kelvin. The QAH states have been proposed for low energy cost electronic devices and topological computation~\cite{qi2010b,wang2015c,lian2018b}, however, such low critical temperature is a major obstacle for practical applications. Seeking high-temperature QAH insulators~\cite{you2019,sunj2020,liy2020,xuan2022,sun2020,li2022} has become an important goal in condensed matter physics and material sciences.

The general mechanism for QAH insulator is the spin polarized band inversion~\cite{liu2008}, where both the ferromagnetic (FM) ordering and spin-orbit coupling (SOC) are sufficiently strong. Physically, the ferromagnetism favors transition metal elements with 3$d$ electron, while strong SOC prefers heavy elements. These considerations constitute the essential ingredients for QAH effect in magnetic TI by introducing magnetic dopants or intercalating magnetic layer~\cite{yu2010,wang2015d,zhang2019,li2019,otrokov2019a}. However, the inhomogeneities from magnetic dopants~\cite{chong2020} and defects~\cite{garnica2022} dramatically suppress the exchange gap by several order of magnitude, which fundamentally limits the fully quantized anomalous Hall effect to very low temperatures. The challenge in searching for high-temperature QAH insulators is to synergize the seemingly conflicting requirements of ferromagnetism and SOC.

Here we propose the monolayer $A$Ti$X$ class (KTiSb, KTiBi, RbTiSb, SrTiSn) are potential candidates for large-gap QAH insulators with high-temperature. Their stability, magnetic, electronic, and topological properties are comprehensively investigated by first principles calculations. Remarkably, these materials have Chern number $\mathcal{C}=2$ in the FM ground state, where topology and magnetism are purely from $d$-orbitals of Ti. We further construct the tight-binding model from orbital projected band structure and symmetry analysis to reveal the origin of topology. Remarkably, the band inversion is due to tetrahedral crystal field and hopping, while SOC only \emph{trivially} gap out the Dirac cone at Fermi level. The general physics from the $3d$ orbitals here applies to material class of transition metal compounds with the space group $P4/nmm$ or $P$-$42m$ and their subgroups.

\begin{figure}[b]
\begin{center}
\includegraphics[width=3.3in, clip=true]{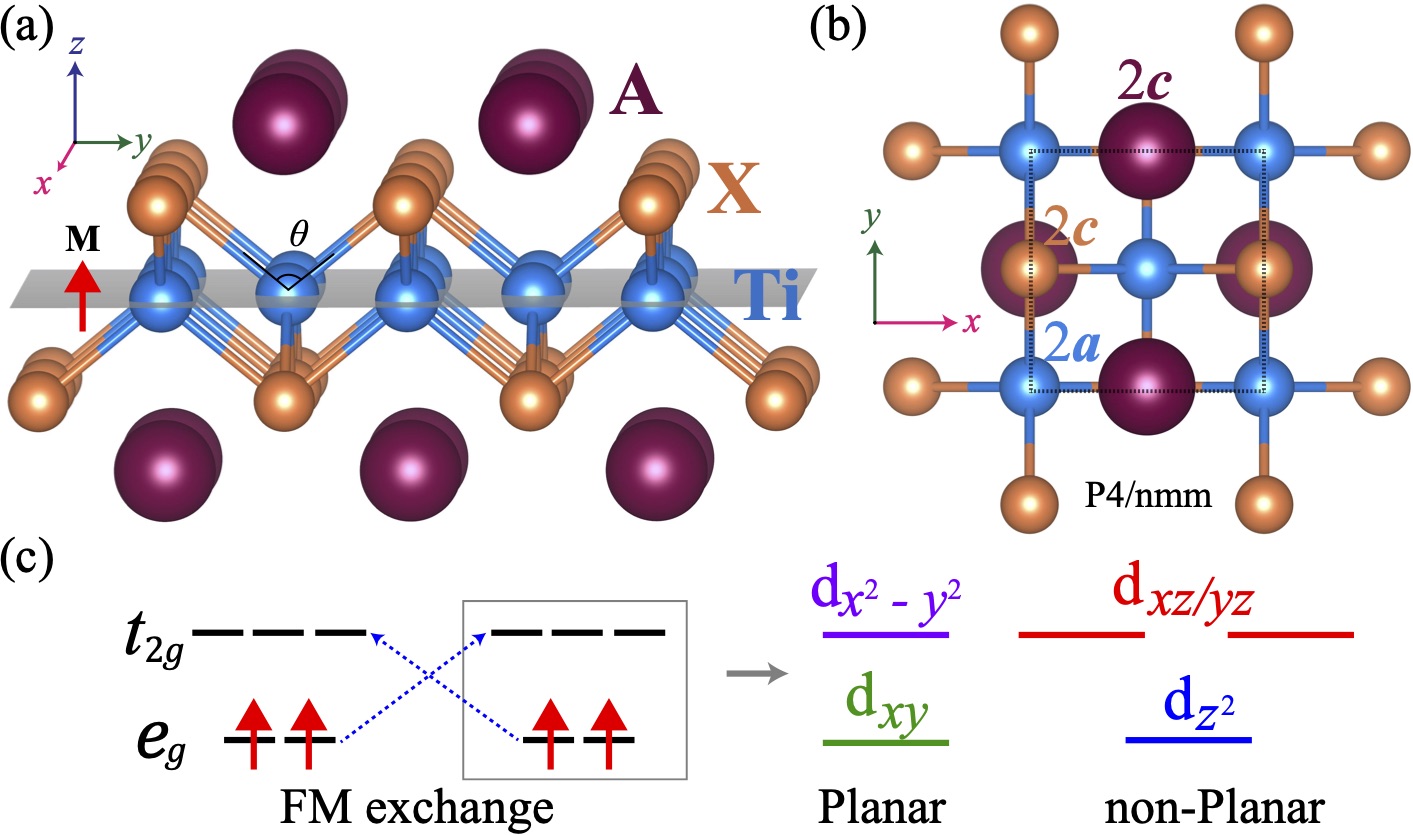}
\end{center}
\caption{(a) and (b) Side and top views of $A$Ti$X$ monolayer. The 2D material class has a FM ground state along $z$ axis with spin magnetic moment $2\mu_B$ per Ti atom. $\theta$ is bond angle of $X$-Ti-$X$. The Wyckoff positions $2a$ and $2c$ are displayed (notation adopted from Bilbao Crystallographic Server~\cite{bilbao2,bilbao3,bilbao1}). The key symmetry operations include $C_{4z}$ rotation, $M_x$ and $M_y$ mirrors, $C_{2x}$ and $C_{2y}$ rotations. (c) Crystal field splitting and schematic diagram of the FM kinetic exchange coupling between Ti atoms.}
\label{fig1}
\end{figure}

{\color{blue}
\emph{Structure and magnetic properties.}}
The monolayer $A$Ti$X$ class has a tetragonal lattice with the space group $P4/nmm$ (No.~129), similar to LiFeSe~\cite{liy2020}. As shown in Fig.~\ref{fig1}(a), each primitive cell includes five atomic layers, where each Ti atom is surrounded by four $X$ atoms forming a distorted edge-sharing tetrahedron, and the group IA (or IIA) element $A$ with an ultralow electronegativity easily loses valence electrons and becomes $A^+$ (or $A^{2+}$). 
These QAH materials are obtained from high-throughput screening of insulating $A$Ti$X$ with $X$ from group IVA and VA. Their lattice constants are listed in Table~\ref{tab1}. The absence of imaginary phonon frequency from the phonon spectra calculations suggest the dynamical stability of monolayer $A$Ti$X$ structure (Fig.~S1~\cite{supple}). We will mainly discuss KTiSb with similar results for this class.

\begin{table}[t]
\caption{Lattice constants, bond angle, MAE per unit cell, and Curie temperature $T_c$ from Monte Carlo simulations.} 
\begin{center}\label{tab1}
\renewcommand{\arraystretch}{1.4}
\begin{tabular*}{3.4in}
{@{\extracolsep{\fill}}ccccc}
\hline
\hline
 & $a$ (\AA) & $\theta$ ($^\circ$) & MAE (meV) & $T_c$ (K)\\
\hline
KTiSb & 4.49 & 101 & 2.6 & 637\\
KTiBi & 4.56 & 99 & 9.8 & 662\\
RbTiSb & 4.53 & 103 & 2.0 & 598\\
SrTiSn & 4.50 & 98 & 1.0 & 436\\
\hline
\hline
\end{tabular*}
\end{center}
\end{table}

To determine the magnetic ground state of monolayer $A$Ti$X$, we compare five magnetic configurations: (i) FM, (ii) checkboard antiferromagnetic (AFM), (iii) collinear AFM, (iv) zigzag AFM, and (v) big zigzag AFM (see Fig.~S2~\cite{supple}), and find that FM is the ground state (Table~S1~\cite{supple}). Take KTiSb for example, the calculated energy of FM state is about $290$, $277$, $146$, and $205$~meV/Ti lower than the four AFM states, respectively. The energy difference is found to be similar for other monolayer $A$Ti$X$. The large magnetic energy difference indicates the FM exchange coupling is strong. The magnetocrystalline anisotropy energy (MAE), defined as the total energy difference between in-plane and out-of- plane spin configurations, are listed in Table~\ref{tab1}, while a positive MAE implies an out-of-plane easy axis.

To elucidate the underlying mechanism of ferromagnetism, we analyze the orbital occupation of Ti atoms. The $d$ orbitals are split by the tetrahedral crystal field into doublet $e_g (d_{xy},d_{z^2})$ and triplet $t_{2g}(d_{x^2-y^2},d_{xz},d_{yz})$ orbitals (Fig.~\ref{fig1}(c) with coordinates in Fig.~\ref{fig1}(b)). The energy of $e_g$ stays lower with respect to $t_{2g}$, because the latter point towards the negatively charged ligands. Each Ti atom is in the $e^2_gt_{2g}^0$ configuration with the magnetic moment of $2\mu_B$ according to the Hund's rule, which is consistent with the density functional theory calculation. The exchange between two neighboring Ti atoms with closed $e_{g}^2$ subshell must be AFM, which is associated with the Pauli exclusion principle in the virtually excited $e_g^1e_g^3$ state. Meanwhile, the superexchange from near $90^\circ$ Ti-$X$-Ti bond leads to extremely weak FM only~\cite{khomskii2004}. The key is rooted in the FM $e_g$-$t_{2g}$ kinetic exchange, the large radius of ligands only cause small crystal field splitting which further make FM $e_g$-$t_{2g}$ exchange dominates over AFM $e_g$-$e_{g}$ exchange. Furthermore, the \emph{ab initio} Heisenberg exchange parameters of KTiSb are $J_1=-72.6$~meV, $J_2=-33.1$~meV, and $J_3=16.4$~meV for nearest-, next-nearest-, and next-nearest-neighbor Ti-Ti pairs, respectively~\cite{supple}. Negative $J_i$ means the FM exchange coupling. The Curie temperature for monolayer $A$Ti$X$ are listed in Table~\ref{tab1} (Fig.~S3) by Monte Carlo simulations~\cite{supple}.

\begin{figure}[t]
\begin{center}
\includegraphics[width=3.4in, clip=true]{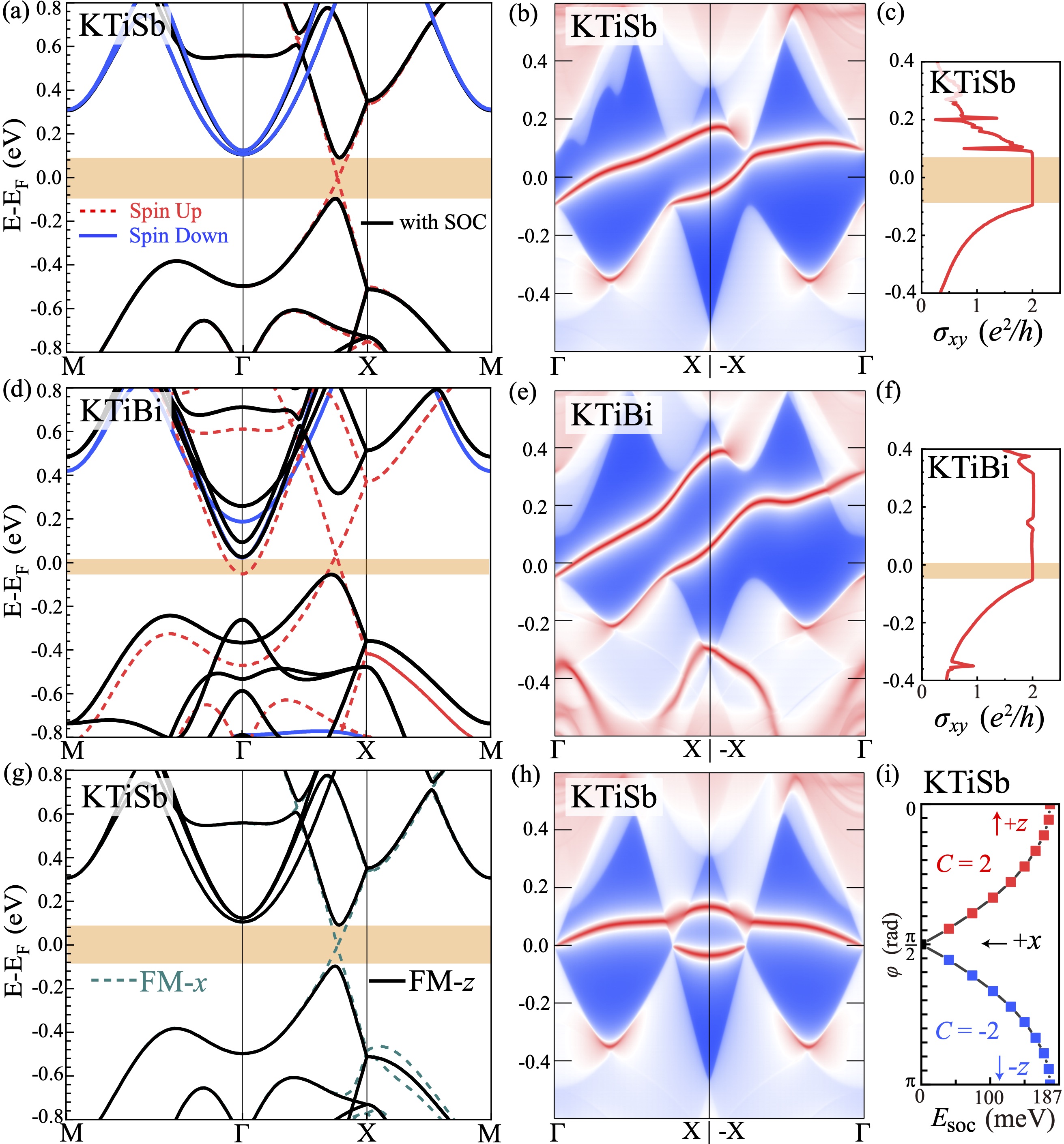}
\end{center}
\caption{Electronic structure and topological properties of monolayer KTiSb and KTiBi. (a)-(c) KTiSb, (d)-(f) KTiSb, The band structure with and without SOC; topological edge states calculated along $x$ direction; anomalous Hall conductance $\sigma_{xy}$ as a function of Fermi energy, respectively. (g) The band structure for FM along $z$- and $x$-axis of KTiSb. (h) The edge states along $x$ axis under FM-$x$ state of KTiSb. (i) Dependence of band gap and $\mathcal{C}$ on the spin orientation quantified by a polar angle $\varphi$, where $\varphi=0,\pi/2,\pi$ denote the $+z,+x,-z$ directions, respectively.}
\label{fig2}
\end{figure}
 
\begin{figure}[t]
\begin{center}
\includegraphics[width=3.4in, clip=true]{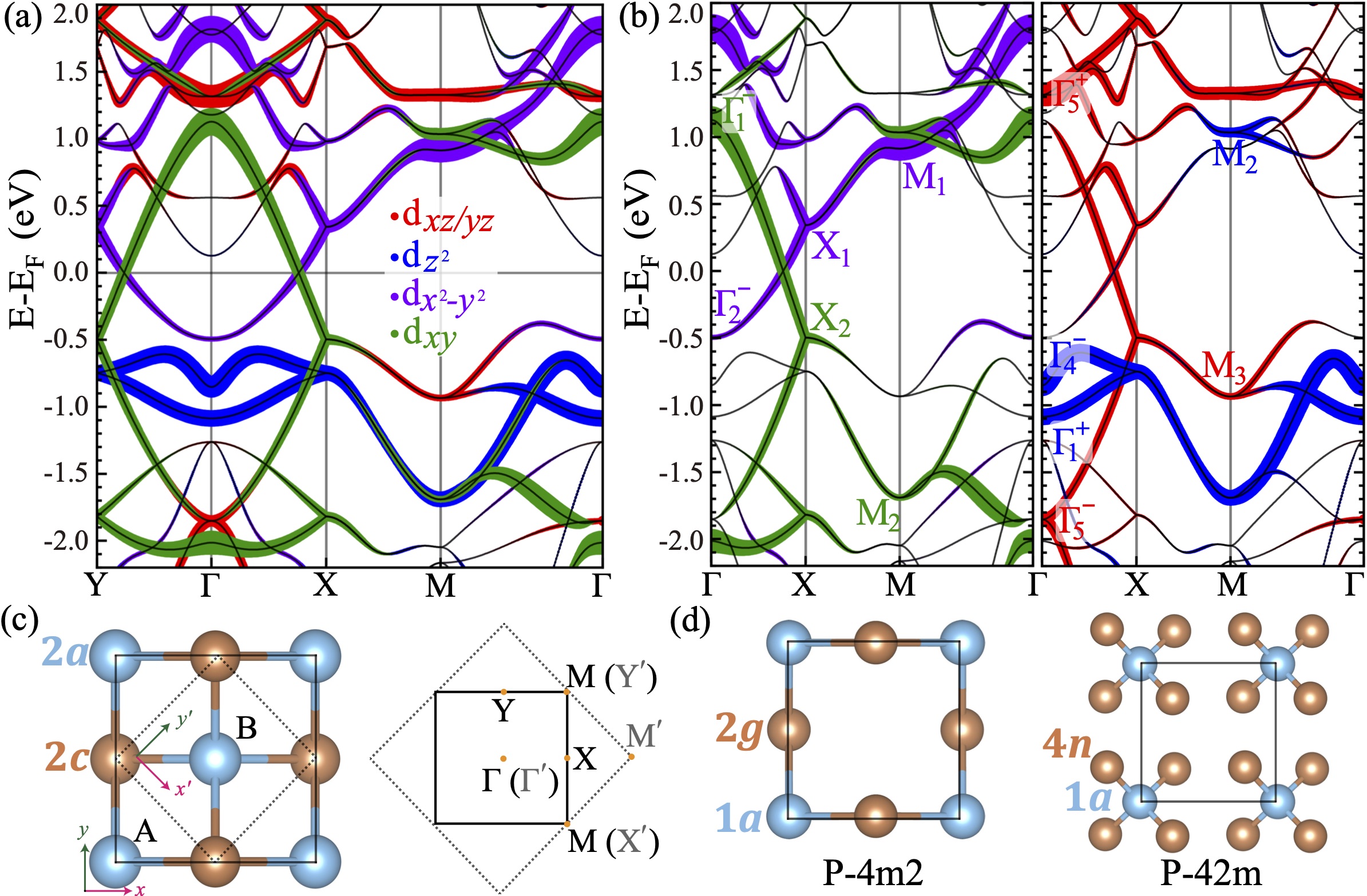}
\end{center}
\caption{(a) The $d$-orbital projected band structures without SOC for spin up of monolayer KTiSb. (b) The IR of high symmetry point at Brillouin zone boundary, where $(d_{xy},d_{x^2-y^2})$ and $(d_{xz/yz},d_{z^2})$ are separately displayed. (c) The relation between original (labelled by $\Gamma$-$X$-$Y$-$M$) and  unfolded Brillouin zone (labelled by $\Gamma^\prime$-$X^\prime$-$Y^\prime$-$M^\prime$). Here $M^\prime$ and $\Gamma^\prime$ are folded into $\Gamma$. The coordinates ($x,y$) and ($x',y'$) are for original and new unit cell, respectively. (d) Lattices of space group $P$-$4m2$ (No.~115) and $P$-$42m$ (No.~111).}
\label{fig3}
\end{figure}

{\color{blue}\emph{Electronic structures.}} 
Fig.~\ref{fig2}(a) displays the electronic structure of monolayer KTiSb with and without SOC. In the absence of SOC, the spin-down bands has an insulating gap, and the spin up bands form a spin polarized Dirac semimetal. Specifically, the two spin-up bands near the Fermi level are mainly contributed by $d_{x^2-y^2}$ and $d_{xy}$ orbitals of Ti (Fig.~\ref{fig3}(a)), where the Dirac points along $\Gamma$-$X$ ($\Gamma$-$Y$) are protected by $M_y$ ($M_x$). When including SOC, the out-of-plane ferromagnetism breaks $M_x$ and $M_y$, and SOC opens a Dirac gap. The anomalous Hall conductance $\sigma_{xy}$ as a function of Fermi energy is calculated in Fig.~\ref{fig2}(c), which displays a quantized value of $2e^2/h$ within the bulk gap of 187~meV. This indicates the topological nontrivial bands with Chern number $\mathcal{C}=2$, which is consistent with two chiral edge states dispersing within the bulk gap as in the edge local density of states calculations (Fig.~\ref{fig2}(b)). By further varying the spin orientation from $-z$ to $+x$, then to the $+z$ axis, the band gap monotonically decreases to close, and then reopens, which is accompanied by the topological phase transitions from $\mathcal{C}=-2$ to $\mathcal{C}=2$ in Fig.~\ref{fig2}(i). For in-plane ferromagnetism along $x$ axis, the gapless Dirac points along $\Gamma$-$X$ (Fig.~\ref{fig2}(g)) are protected by $C_{2x}$, while the Dirac points along $\Gamma$-$Y$ are protected by $M_x$. Electronic structure of monolayer KTiBi with similar topological properties are shown in Fig.~\ref{fig2}(d)-\ref{fig2}(f), with an indirect band gap of 80~meV. The sizable band gap (Table~S2) in this family is attributed to the enhanced effective SOC strength of Ti-$3d$ orbitals by bonding with the ligand heavy elements.

{\color{blue}\emph{Symmetry and origin of topology.}}
To reveal the origin of $\mathcal{C}=2$ topology in the electronic structure, we perform a systematic investigation of orbital projected band structure and symmetry analysis of irreducible representations (IR), and further construct a tight-binding model to recover the essential topological physics. Remarkably, different from the conventional $s$-$p$ or $s$-$d$ band inversion, where SOC not only induces topological band inversion but also opens the Dirac gap. Here, the band inversion is due to crystal field and hopping, while SOC only trivially gap out the Dirac cone at Fermi level.

\begin{figure}[t]
\begin{center}
\includegraphics[width=3.4in, clip=true]{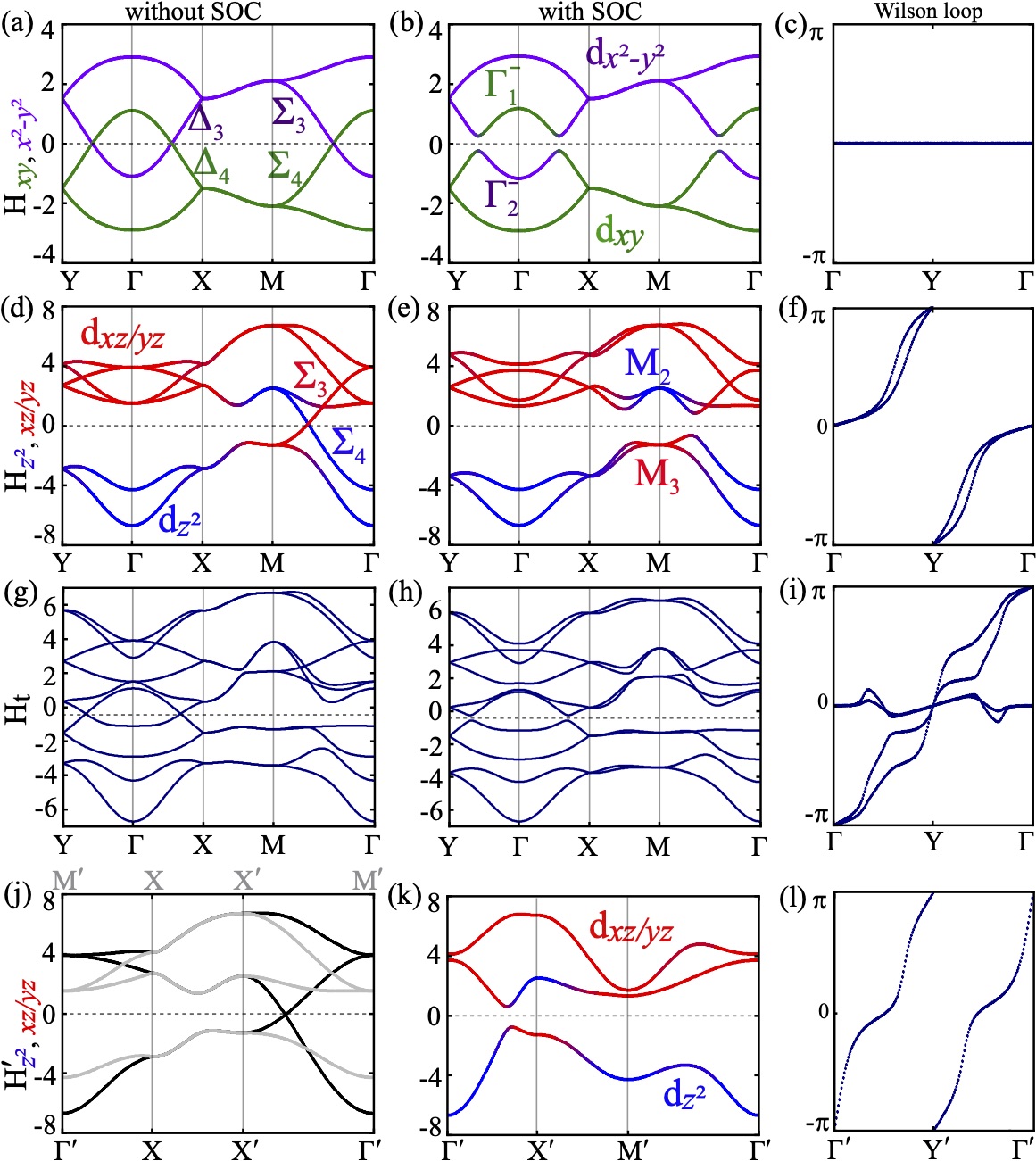}
\end{center}
\caption{The band structure and Wilson loop of the tight-binding model. (a)-(c) $\mathcal{H}_1$, (d)-(f) $\mathcal{H}_2$, (g)-(i) $\mathcal{H}$ from all five $d$ orbitals. The band structure without SOC, with SOC and Wilson loop of the lowest bands below Fermi level, respectively. The IR along symmetry lines and at symmetry points in (a) and (d) are consistent with that in Fig.~\ref{fig3}(b). (j)-(l) The band structure of $\mathcal{H}_2^\prime$ from $(d_{xz},d_{yz},d_{z^2})$ in the new unit cell with SOC, without SOC and Wilson loop of the lowest band, repsectively.}
\label{fig4}
\end{figure}

The orbital projected (spin up) band structures without SOC is shown in Fig.~\ref{fig3}(a). The band inversion between $d_{xy}$ and $d_{x^2-y^2}$ at Fermi level induces four Dirac points, which are gapped by SOC. Naively, these features seem to explain $\mathcal{C}=2$ phase well with each gapped Dirac point contributing $\mathcal{C}=1/2$. However, by calculating IR of the inverted bands at $\Gamma$ (Fig.~\ref{fig3}(b) left), we find that IR of the two bands $\Gamma^-_1$ and $\Gamma^-_2$ have the same $C_{4z}$ eigenvalue $1$ (see character table in Table~S3), which indicate the band inversion between $d_{xy}$ and $d_{x^2-y^2}$ is trivial and does not contribute Chern number. We further write down a tight-binding model from the \emph{planar} orbitals $(d_{xy},d_{x^2-y^2})$ as $\mathcal{H}_1=\sum_{\langle ij\rangle}[d^\dag_{i}\hat{t}_{ij}d_j+\text{H.c.}]+\sum_{\langle\langle ij\rangle\rangle}[d^\dag_{i}\hat{t}^\prime_{ij}d_j+\text{H.c.}]$, where $\langle ij\rangle$ and $\langle\langle ij\rangle\rangle$ denote the nearest-neighbor (NN) and next-nearest-neighbor (NNN) sites, respectively, $d\equiv(d^A_{xy},d^A_{x^2-y^2},d^B_{xy},d^B_{x^2-y^2})^T$, $A$ and $B$ denote the sublattice, $\hat{t}_{ij}$ and $\hat{t}_{ij}'$ are hopping matrices terms with SOC included~\cite{supple}. As displayed in Fig.~\ref{fig4}(a)-\ref{fig4}(c), the Dirac points around $\Gamma$ from the band inversion are indeed gapped by SOC, while the Wilson loop calculation confirms the trivial topology consistent with symmetry consideration.  Furthermore, as indicated by IR shown in Fig.~\ref{fig4}(a), the band structure of $\mathcal{H}_1$ without SOC has level crossing along $\Gamma$-$M$, which is contrary to the first principles calculations. All of these results suggest the topology must originate from other $d$ orbitals.

Another band inversion occurs at $M$ between $d_{xz}, d_{yz}$ and $d_{z^2}$ (Fig.~\ref{fig3}(b) right),  which is about $1$~eV away from the Fermi level. The calculated IR are consistent with the Elementary Band Representations (EBR)~\cite{slager2017,bradlyn2017,vergniory2017,elcoro2017,bouhon2021} listed in Table~\ref{tab2}. We further construct a concrete tight-binding model $\mathcal{H}_2$ from \emph{nonplanar} orbitals $d_{xz}, d_{yz}$ and $d_{z^2}$ to decipher whether the band inversion at $M$ contributes nontrivial topology. $\mathcal{H}_2=\sum_{\langle ij\rangle}[c^\dag_{i}\hat{h}_{ij}c_j+\text{H.c.}]+\sum_{\langle\langle ij\rangle\rangle}[c^\dag_{i}\hat{h}^\prime_{ij}c_j+\text{H.c.}]$, where $c\equiv(d^A_{xz},d^A_{yz},d^A_{z^2},d^B_{xz},d^B_{yz},d^B_{z^2})^T$, and $\hat{h}_{ij}$ and $\hat{h}_{ij}'$ are general $6\times 6$ matrices denoting hopping terms with SOC included, which can be simplified by symmetry considerations. The explicit forms are listed in Supplementary Materials~\cite{supple}. As displayed in Fig.~\ref{fig4}(d)-\ref{fig4}(f), we use typical parameters by matching all of the IR along symmetry lines and symmetry points with Fig.~\ref{fig3}(b) and recover band inversion between $M_2$ and $M_3$. The four Dirac points along $\Gamma$-$M$ gapped by SOC contribute a total Chern number $\mathcal{C}=4\times1/2=2$, which is consistent with Wilson loop of the two lowest bands. Thus the band inversion at $M$ accounts for $\mathcal{C}=2$ in monolayer $A$Ti$X$. 

\begin{table}[b]
\caption{EBR without time-reversal symmetry for space group $P4/nmm$ from Wyckoff position $2a$.} 
\begin{center}\label{tab2}
\renewcommand{\arraystretch}{1.4}
\begin{tabular*}{3.4in}
{@{\extracolsep{\fill}}cccc}
\hline
\hline
   & $\Gamma$ & $X$ & $M$ 
\\ 
\hline
$d_{xy}$ & $\Gamma_1^-(1)\oplus \Gamma_4^+(1)$ & $X_2(2)$ & $M_2(2)$
\\
$d_{x^2-y^2}$ & $\Gamma_2^-(1)\oplus \Gamma_3^+(1)$ & $X_1(2)$ & $M_1(2)$
\\
$d_{z^2}$ & $\Gamma_1^+(1)\oplus \Gamma_4^-(1)$ & $X_1(2)$ & $M_2(2)$
\\
$d_{xz},d_{yz}$ & $\Gamma_5^+(2)\oplus \Gamma_5^-(2)$ & $X_1(2)\oplus X_2(2)$ & $M_3(2)\oplus M_4(2)$
\\
\hline
\hline
\end{tabular*}
\end{center}
\end{table}

To fully understand the band structure and topology in Fig.~\ref{fig3}(a), we need to include all of the five $d$ orbitals. From the above analysis, $\mathcal{H}_1$ from ($d_{xy},d_{x^2-y^2}$) without SOC accounts for the trivial band crossing along $\Gamma$-$X$ and $\Gamma$-$Y$, but it inevitably introduces level crossing along $\Gamma$-$M$. $\mathcal{H}_2$ from ($d_{xz},d_{yx},d_{z^2}$) without SOC accounts for the topological band inversion at $M$, but it inevitably introduces level crossing along $\Gamma$-$M$. As shown in Fig.~\ref{fig4}(a) and \ref{fig4}(d), the IRs of the two crossed bands along $\Gamma$-$M$ are $\Sigma_3$ and $\Sigma_4$ in $\mathcal{H}_1$, which is just the opposite to that in $\mathcal{H}_2$. This symmetry analysis simply indicates by including inter-orbital hopping (as $\mathcal{H}_{12}$) between $(d_{xy},d_{x^2-y^2})$ and ($d_{xz},d_{yx},d_{z^2}$), the symmetry protected band crossing along $\Gamma$-$M$ will be gapped even in the absence of SOC. This is clearly demonstrated in Fig.~\ref{fig4}(g) for $\mathcal{H}=\mathcal{H}_1+\mathcal{H}_2+\mathcal{H}_{12}$ without SOC~\cite{supple}, which leaves the Dirac points along $\Gamma$-$X$ and $\Gamma$-$Y$ only. Then putting the SOC back, we see a full gap opening in Fig.~\ref{fig4}(h), and the Wilson loop for the lowest four band is topological which is equal to the direct sum of Fig.~\ref{fig4}(c) and Fig.~\ref{fig4}(f). There is a subtlety point in Fig.~\ref{fig3}(a) worth mentioning, that at $\Gamma$ the two-fold degenerate IR $\Gamma^-_5(2)$ from $d_{xz},d_{yz}$ energetically lies below $\Gamma_1^+(1)\oplus\Gamma_4^-(1)$ from $d_{z^2}$. This will not affect the topology due to opposite angular momentum from $d_{xz}\pm id_{yz}$. Now, we understand the essential physics and reproduce the main result in Fig.~\ref{fig3}(a)and \ref{fig3}(b) by a five $d$-orbital model in Fig.~\ref{fig4}(h). Since the topological band inversion is contributed by $d_{xz}$, $d_{yx}$, and $d_{z^2}$ far away from Fermi level, the chiral edge states alway exist irrespective of the band crossing or gap opening along $\Gamma$-$X$ ($Y$). This explains the edge states in Fig.~\ref{fig2}(h) for FM along $x$ axis.

Finally, we provide an intuitive understanding why the band inversion from $(d_{xz},d_{yz})$ and $d_{z^2}$ at $M$ leads to high Chern number $\mathcal{C}=2$. For the unit cell in Fig.~\ref{fig1}(b), the elements at Wyckoff position $2c$ introduce the difference between A and B sublattice of Ti at Wyckoff position $2a$. Since the topology is purely from the $d$ orbitals of Ti, if we ignore the difference between $A$ and $B$ sublattices, we can construct a new square lattice with the unit cell containing a single Ti atom (gray dash line in Fig.~\ref{fig3}(c)). The relation between unfolded (labelled by $\Gamma^\prime$-$X^\prime$-$Y^\prime$-$M^\prime$) and original Brillouin zone is also displayed. Now we can rewrite $\mathcal{H}_2$ in the new unit cell as $\mathcal{H}_2^\prime$ from nonplanar orbitals $d_{xz},d_{yz}$ and $d_{z^2}$ only~\cite{supple}, and replot the band structure. As shown in Fig.~\ref{fig4}(j), the band structure without SOC along $\Gamma^\prime$-$X$-$X^\prime$-$\Gamma^\prime$ (black line) together with $M^\prime$-$X$-$X^\prime$-$M^\prime$ (gray line) almost reproduces the band features in Fig.~\ref{fig4}(d), i.e., the band structure of $\mathcal{H}_1$ is almost the folding of that of $\mathcal{H}_1^\prime$. According to band folding picture, the band inversion at $M$ in the original Brillouin zone now is related to band inversion at $X^\prime$ in the unfolded Brillouin zone as shown in Fig.~\ref{fig4}(k). There is $C_{2z}$ rotational symmetry at $X^\prime$, so the band inversion between angular momentum $\ell_z=\pm 1$ of $d_{xz},d_{yz}$ and $\ell_z=0$ of $d_{z^2}$ at $X^\prime$ is similar to a $s$-$p$ band inversion with $\mathcal{C}=1$. The same band inversion also happens around $Y^\prime$ from $C_{4z}$ with another $\mathcal{C}=1$. Then the total Chern number is $\mathcal{C}=1+1=2$, which is consistent with the Wilson loop calculation in Fig.~\ref{fig4}(l). Now both $X'$ and $Y'$ are folded into $M$ without gap closing, then the total Chern number of the folded bands below Fermi level remains the same.

{\color{blue}\emph{Generalization and discussion}.}
The model and analysis from $d$ orbitals above are general, and also apply to the previous reported $\mathcal{C}=2$ QAH insulator in monolayer TiTe~\cite{xuan2022}, LiFeSe~\cite{liy2020}, FeI~\cite{sun2020} with similar $d$-orbital projected band structure and IR (Fig.~S9~\cite{supple}), where the origin of topology from $(d_{xz},d_{yz})$ and $d_{z^2}$ at $M$ were overlooked. The topological physics in TiTe is from $d$-orbitals of Ti at Wyckoff position $2a$ of space group $P4/nmm$, which is in the $e^{\uparrow2}_gt_{2g}^{\uparrow0}$ configuration with occupied $d_{xy}$ and $d_{z^2}$ for majority spin in the polarized state. While in LiFeSe~\cite{liy2020} and FeI~\cite{sun2020}, each Fe takes the Wyckoff position $2a$ of $P4/nmm$ and is in the $e^4_gt_{2g}^3=e^{\uparrow2}_gt_{2g}^{\uparrow3}e^{\downarrow2}_gt_{2g}^{\downarrow0}$ configuration. The $e_g$-$t_{2g}$ kinetic exchange leads to ultrastable FM, and the system is in a spin polarized state with occupied $d_{xy}$ and $d_{z^2}$ for minority spin. The similar polarized $e^{2}_gt_{2g}^{0}$ configuration with $C_{4z}$ rotational symmetry in all of these materials (Ti or Fe lattices) give rise to $\mathcal{C}=2$ QAH phase with the same origin of topology.

\begin{table}[t]
\caption{Two typical layer groups and their subgroups which are compatible with distorted tetrahedral crystal field and $C_{4z}$ rotational symmetry.} 
\begin{center}\label{tab3}
\renewcommand{\arraystretch}{1.4}
\begin{tabular*}{3.4in}
{@{\extracolsep{\fill}}ccc}
\hline
\hline
Space group & $P4/nmm$ & $P$-$42m$ 
\\
Subgroup & $P4/n$, $P42_12$, $P$-$4m2$, $P$-$42_1m$ & N/A
\\
\hline
\hline
\end{tabular*}
\end{center}
\end{table}

The key ingredient for the $\mathcal{C}=2$ phase is the polarized $e^{2}_gt_{2g}^{0}$ configuration of $d$ orbitals, which is generated by the distorted tetrahedral crystal field by ligands at Wyckoff position $2c$. The distortion of tetrahedron will lift the degeneracy within $t_{2g}$ and $e_g$, and determine the splitting and relative energetics between planar and nonplanar orbitals (Fig.~S10~\cite{supple}). As long as the crystal field splitting is small compared to the $d$ orbital band width, the five $d$ orbital model here always works, namely, $d_{xz},d_{yz}$ and $d_{z^2}$ give rise to the topological band inversion. Here we suggest several layer groups in Table~\ref{tab3}, which are compatible with such a crystal field and have square lattice structure with $C_{4z}$ symmetry. For example, the most simple structures are $P$-$4m2$ and $P$-$42m$ shown in Fig.~\ref{fig3}(d). We propose similar electronic structure with same topological properties can be found in transition metal compounds with these space groups, when the $d$ orbital configuration $e^{2}_gt_{2g}^{0}$ is fulfilled. Interestingly, the octahedral crystal field could also lead to $t_{2g}$ and $e_g$ splitting with opposite energy sequence, the polarized $t_{2g}^{3}e^{0}_g$ configuration of $d$ orbitals together with $C_{4z}$ from certain transition metal may also lead to $\mathcal{C}=2$ QAH phase, the concrete material family is left for future study. 

In summary, our work uncover $\mathcal{C}=2$ QAH phase from $3d$ orbitals which applies to a large class of materials in the space group $P4/nmm$. We hope the theoretical work here can aid the search for new QAH insulators in transition metal compounds.

\begin{acknowledgments}
{\color{blue}\emph{Acknowledgment.}} This work is supported by the National Key Research Program of China under Grant No.~2019YFA0308404, the Natural Science Foundation of China through Grant No.~12174066, Science and Technology Commission of Shanghai Municipality under Grant No.~20JC1415900, the Innovation Program for Quantum Science and Technology through Grant No.~2021ZD0302600, Shanghai Municipal Science and Technology Major Project under Grant No.~2019SHZDZX01.
\end{acknowledgments}

\end{document}